%% file: main.tex
\documentclass{article}
\input{preamble}

\title{WaveGlow: A Flow-based Generative Network for Speech Synthesis}
\name{Ryan Prenger, Rafael Valle, Bryan Catanzaro}
\address{NVIDIA Corporation}

\begin{document}
\maketitle

\begin{abstract}
In this paper we propose WaveGlow: a flow-based network capable of generating high quality speech from mel-spectrograms. WaveGlow combines insights from Glow~\cite{kingma2018glow} and WaveNet~\cite{van2016wavenet} in order to provide fast, efficient and high-quality audio synthesis, without the need for auto-regression.  WaveGlow is implemented using only a single network, trained using only a single cost function: maximizing the likelihood of the training data, which makes the training procedure simple and stable.  Our PyTorch implementation produces audio samples at a rate of more than 500 kHz on an NVIDIA V100 GPU. Mean Opinion Scores show that it delivers audio quality as good as the best publicly available WaveNet implementation. All code will be made publicly available online~\cite{WaveGlowSite}.
\end{abstract}


\section{Introduction} \label{sec:introduction}
\input{introduction}
\section{WaveGlow}\label{sec:audio_flow}
\input{audio_flow}
\section{Experiments}\label{sec:experiments}
\input{experiments}
\section{Discussion}\label{sec:discussion}
\input{discussion}
\subsubsection*{Acknowledgments}
\input{acknowledgments}

\newpage
\bibliographystyle{IEEEbib}
\bibliography{main}
\end{document}

%% file: preamble.tex
\usepackage{hyperref} 
\usepackage{spconf,amsmath,graphicx}
\usepackage{pgfplots}
\usepackage{siunitx} 
\usepackage{caption}
\pgfplotsset{compat=1.14}

%% file: introduction.tex
As voice interactions with machines become increasingly useful, efficiently synthesizing high quality speech becomes increasingly important.  Small changes in voice quality or latency have large impacts on customer experience and customer preferences.  However, high quality, real-time speech synthesis remains a challenging task. Speech synthesis requires generating very high dimensional samples with strong long term dependencies.  Additionally, humans are sensitive to statistical imperfections in audio samples.  Beyond the quality challenges, real-time speech synthesis has challenging speed and computation constraints. Perceived speech quality drops significantly when the audio sampling rate is less than 16kHz, and higher sampling rates generate even higher quality speech. Furthermore, many applications require synthesis rates much faster than 16kHz.  For example, when synthesizing speech on remote servers, strict interactivity requirements mean the utterances must be synthesized quickly at sample rates far exceeding real-time requirements.

Currently, state of the art speech synthesis models are based on parametric neural networks.  Text-to-speech synthesis is typically done in two steps. The first step transforms the text into time-aligned features, such as a mel-spectrogram~\cite{wang2017tacotron,shen2017natural}, or $\mathrm{F0}$ frequencies and other linguistic features~\cite{van2016wavenet,arik2017deep}. A second model transforms these time-aligned features into audio samples.  This second model, sometimes referred to as a vocoder, is computationally challenging and affects quality as well. We focus on this second model in this work.  Most of the neural network based models for speech synthesis are auto-regressive, meaning that they condition future audio samples on previous samples in order to model long term dependencies.  These approaches are relatively simple to implement and train.  However, they are inherently serial, and hence can't fully utilize parallel processors like GPUs or TPUs.  Models in this group often have difficulty synthesizing audio faster than 16kHz without sacrificing quality.

At this time we know of three neural network based models that can synthesize speech without auto-regression: Parallel WaveNet~\cite{van2016wavenet}, Clarinet~\cite{ping2018clarinet}, and MCNN for spectrogram inversion~\cite{arik2018fast}. These techniques can synthesize audio at more than 500kHz on a GPU.  However, these models are more difficult to train and implement than the auto-regressive models.  All three require compound loss functions to improve audio quality or problems with mode collapse~\cite{van2018parallel,ping2018clarinet,arik2018fast}.  In addition, Parallel WaveNet and Clarinet require two networks, a student network and teacher network.  The student networks underlying both Parallel WaveNet and Clarinet use Inverse Auto-regressive Flows (IAF)~\cite{kingma2016improved}. Though the IAF networks can be run in parallel at inference time, the auto-regressive nature of the flow itself makes calculation of the IAF inefficient. To overcome this, these works use a teacher network to train a student network on a approximation to the true likelihood. These approaches are hard to reproduce and deploy because of the difficulty of training these models successfully to convergence.

In this work, we show that an auto-regressive flow is unnecessary for synthesizing speech.  Our contribution is a flow-based network capable of generating high quality speech from mel-spectrograms. We refer to this network as WaveGlow, as it combines ideas from  Glow~\cite{kingma2018glow} and WaveNet~\cite{van2016wavenet}.  WaveGlow is simple to implement and train, using only a single network, trained using only the likelihood loss function.  Despite the simplicity of the model, our PyTorch implementation synthesizes speech at more than 500kHz on an NVIDIA V100 GPU: more than 25 times faster than real time. Mean Opinion Scores show that it delivers audio quality as good as the best publicly available WaveNet implementation trained on the same dataset.

%% file: audio_flow.tex
WaveGlow is a generative model that generates audio by sampling from a distribution. To use a neural network as a generative model, we take samples from a simple distribution, in our case, a zero mean spherical Gaussian with the same number of dimensions as our desired output, and put those samples through a series of layers that transforms the simple distribution to one which has the desired distribution. In this case, we model the distribution of audio samples conditioned on a mel-spectrogram. 

\begin{gather}
\boldsymbol{z} \sim \mathcal{N}(\boldsymbol{z};0,\boldsymbol{I}) \\
\boldsymbol{x} = \boldsymbol{f}_0 \circ \boldsymbol{f}_1 \circ \ldots \boldsymbol{f}_k(\boldsymbol{z})
\end{gather}

We would like to train this model by directly minimizing the negative log-likelihood of the data. If we use an arbitrary neural network this is intractable.  Flow-based networks~\cite{dinh2014nice,dinh2016density,kingma2018glow} solve this problem by ensuring the neural network mapping is invertible.  By restricting each layer to be bijective, the likelihood can be calculated directly using a change of variables:

\begin{gather}
\log{p_\theta(\boldsymbol{x})} = \log{p_\theta(\boldsymbol{z})} + \sum_{i=1}^{k} \log
|\det(\boldsymbol{J}(\boldsymbol{f}_i^{-1}(\boldsymbol{x})))| \\
\boldsymbol{z} = \boldsymbol{f}_k^{-1} \circ \boldsymbol{f}_{k-1}^{-1} \circ \ldots \boldsymbol{f}_0^{-1}(\boldsymbol{x})
\end{gather}

In our case, the first term is the log-likelihood of the spherical Gaussian.  This term penalizes the $l_2$ norm of the transformed sample.  The second term arises from the change of variables, and the $\boldsymbol{J}$ is the Jacobian. The log-determinant of the Jacobian rewards any layer for increasing the volume of the space during the forward pass.  This term also keeps a layer from just multiplying the $\boldsymbol{x}$ terms by zero to optimize the $l_2$ norm.  The sequence of transformations is also referred to as a normalizing flow~\cite{rezende2015variational}.

\begin{figure}[!ht]
    \centering
    \includegraphics[width=0.35\textwidth]{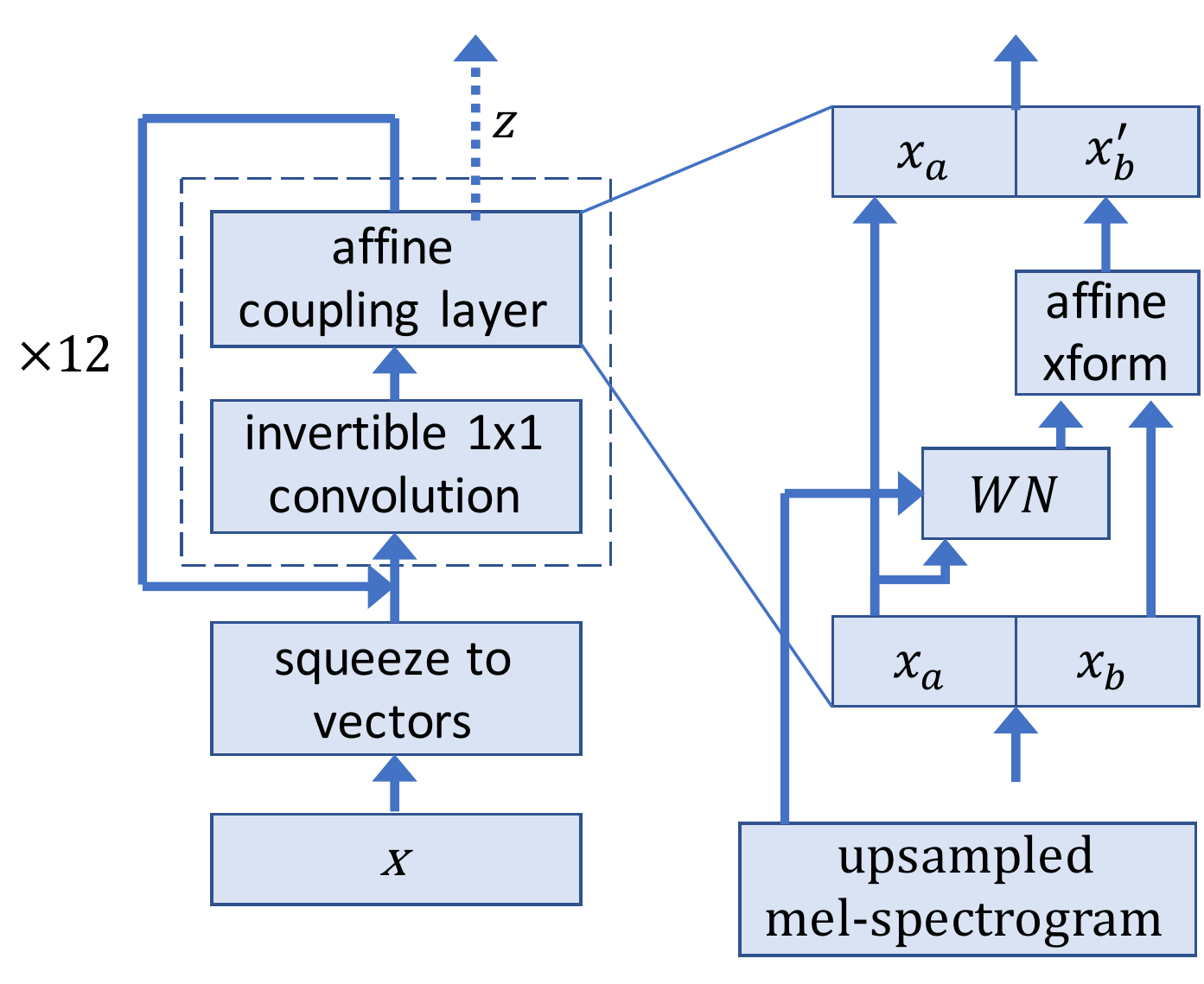}
    \caption{WaveGlow network}
    \label{fig:network_diagram}
\end{figure}

Our model is most similar to the recent Glow work~\cite{kingma2018glow}, and is depicted in figure~\ref{fig:network_diagram}. For the forward pass through the network, we take groups of 8 audio samples as vectors, which we call the "squeeze" operation, as in~\cite{kingma2018glow}. We then process these vectors through several "steps of flow". A step of flow here consists of an invertible $1\times 1$ convolution followed by an affine coupling layer, described below.

\subsection{Affine Coupling Layer}
Invertible neural networks are typically constructed using coupling layers~\cite{dinh2014nice, dinh2016density, kingma2018glow}.  In our case, we use an affine coupling layer~\cite{dinh2016density}.  Half of the channels serve as inputs, which then produce multiplicative and additive terms that are used to scale and translate the remaining channels:  

\begin{gather}
\boldsymbol{x}_a, \boldsymbol{x}_b =  split(\boldsymbol{x}) \\
(\log \boldsymbol{s}, \boldsymbol{t}) = WN(\boldsymbol{x}_a, \mathit{mel\textnormal{-}spectrogram}) \\
\boldsymbol{x}_b\prime = \boldsymbol{s} \odot \boldsymbol{x}_b + \boldsymbol{t} \\
\boldsymbol{f}_{coupling}^{-1}(\boldsymbol{x}) = concat(\boldsymbol{x}_a, \boldsymbol{x}_b\prime)
\end{gather}

Here $WN()$ can be any transformation.  The coupling layer preserves invertibility for the overall network, even though $WN()$ does not need to be invertible. This follows because the channels used as the inputs to $WN()$, in this case $\boldsymbol{x}_a$, are passed through unchanged to the output of the layer. Accordingly, when inverting the network, we can compute $\boldsymbol{s}$ and $\boldsymbol{t}$ from the output $\boldsymbol{x}_a$, and then invert $\boldsymbol{x}_b\prime$ to compute $\boldsymbol{x}_b$, by simply recomputing $WN(\boldsymbol{x}_a, \mathit{mel\textnormal{-}spectrogram})$. In our case, $WN()$ uses layers of dilated convolutions with gated-$\tanh$ nonlinearities, as well as residual connections and skip connections.  This $WN$ architecture is similar to WaveNet~\cite{van2016wavenet} and Parallel WaveNet~\cite{van2018parallel}, but our convolutions have 3 taps and are not causal.  The affine coupling layer is also where we include the mel-spectrogram in order to condition the generated result on the input.  The upsampled mel-spectrograms are added before the gated-$\tanh$ nonlinearites of each layer as in WaveNet~\cite{van2016wavenet}.

With an affine coupling layer, only the $\boldsymbol{s}$ term changes the volume of the mapping and adds a change of variables term to the loss. This term also serves to penalize the model for non-invertible affine mappings.
\begin{equation}
\log |\det(\boldsymbol{J}(\boldsymbol{f}_{coupling}^{-1}(\boldsymbol{x})))| = \log |\boldsymbol{s}|
\end{equation}

\subsection{1x1 Invertible Convolution}
In the affine coupling layer, channels in the same half never directly modify one another.  Without mixing information across channels, this would be a severe restriction. Following Glow~\cite{kingma2018glow}, we mix information across channels by adding an invertible 1x1 convolution layer before each affine coupling layer.  The $W$ weights of these convolutions are initialized to be orthonormal and hence invertible. The log-determinant of the Jacobian of this transformation joins the loss function due to the change of variables, and also serves to keep these convolutions invertible as the network is trained.

\begin{gather}
    \boldsymbol{f}_{conv}^{-1} = \boldsymbol{W} \boldsymbol{x} \\
    \log |\det(\boldsymbol{J}(\boldsymbol{f}_{conv}^{-1}(\boldsymbol{x})))| = \log |\det{\boldsymbol{W}}|
\end{gather}

After adding all the terms from the coupling layers, the final likelihood becomes:
\begin{equation}
\begin{split}
    \log{p_\theta(\boldsymbol{x})} =& -\frac{\boldsymbol{z}(\boldsymbol{x})^{T}\boldsymbol{z}(\boldsymbol{x})}{2\sigma^2} \\
    &+ \sum_{j=0}^{\#coupling}\log \boldsymbol{s}_j(\boldsymbol{x},\mathit{mel\textnormal{-}spectrogram}) \\
    &+ \sum_{k=0}^{\#conv} \log \det |\boldsymbol{W}_k|
\end{split}
\end{equation}

Where the first term comes from the log-likelihood of a spherical Gaussian.  The $\sigma^2$ term is the assumed variance of the Gaussian distribution, and the remaining terms account for the change of variables.

\subsection{Early outputs}
Rather than having all channels go through all the layers, we found it useful to output 2 of the channels to the loss function after every 4 coupling layers.  After going through all the layers of the network, the final vectors are concatenated with all of the previously output channels to make the final $\boldsymbol{z}$.  Outputting some dimensions early makes it easier for the network to add information at multiple time scales, and helps gradients propagate to earlier layers, much like skip connections.  This approach is similar to the multi-scale architecture used in~\cite{kingma2018glow,dinh2016density}, though we do not add additional squeeze operations, so vectors get shorter throughout the network.  

\subsection{Inference}
Once the network is trained, doing inference is simply a matter of randomly sampling $\boldsymbol{z}$ values from a Gaussian and running them through the network. As suggested in~\cite{kingma2018glow}, and earlier work on likelihood-based generative models~\cite{parmar2018image}, we found that sampling $\boldsymbol{z}$s from a Gaussian with a lower standard deviation from that assumed during training resulted in slightxly quality higher audio. During training we used $\sigma=\sqrt{0.5}$, and during inference we sampled $\boldsymbol{z}$s from a Gaussian with standard deviation 0.6.  Inverting the 1x1 convolutions is just a matter of inverting the weight matrices.  The inverse is guaranteed by the loss.  The mel-spectrograms are included at each of the coupling layers as before, but now the affine transforms are inverted, and these inverses are also guaranteed by the loss.
\begin{equation}
\boldsymbol{x}_a = \frac{\boldsymbol{x}_a\prime - \boldsymbol{t}}{\boldsymbol{s}}
\end{equation}

%% file: experiments.tex
For all the experiments we trained on the LJ speech data~\cite{ito2017lj}.  This data set consists of 13,100 short audio clips of a single speaker reading passages from 7 non-fiction books. The data consists of roughly 24 hours of speech data recorded on a MacBook Pro using its built-in microphone in a home environment.  We use a sampling rate of 22,050kHz.  

We use the mel-spectrogram of the original audio as the input to the WaveNet and WaveGlow networks. For WaveGlow, we use mel-spectrograms with 80 bins using $\mathrm{librosa}$ mel filter defaults, i.e. each bin is normalized by the filter length and the scale is the same as HTK. The parameters of the mel-spectrograms are FFT size 1024, hop size 256, and window size 1024.  

\subsection{Griffin-Lim}
As baseline for mean opinion score we compare the popular Griffin-Lim algorithm~\cite{griffin1984signal}.  Griffin-Lim takes the entire spectrogram (rather than the reduced mel-spectrogram) and iteratively estimates the missing phase information by repeatedly converting between frequency and time domain.  For our experiments we use 60 iterations from frequency to time domain. 

\subsection{WaveNet}
We compare against the popular open source WaveNet implementation~\cite{ryuichiwn}.  The network has 24 layers, 4 dilation doubling cycles, and uses 512/512/256, for number of residual, gating, and skip channels respectively.  The network upsamples the mel-spectrogram to full time resolution using 4 separate upsampling layers.  The network was trained for \num{1e6} iterations using the Adam optimizer~\cite{kingma2014adam}. The mel-spectrogram for this network is still 80 dimensions but was processed slightly differently from the mel-spectrogram we used in the WaveGlow network.  Qualitatively, we did not find these differences had an audible effect when changed in the WaveGlow network.  The full list of hyperparameters is available online.

\subsection{WaveGlow}
The WaveGlow network we use has 12 coupling layers and 12 invertible 1x1 convolutions.  The coupling layer networks ($WN$) each have 8 layers of dilated convolutions as described in Section~\ref{sec:audio_flow}, with 512 channels used as residual connections and 256 channels in the skip connections.  We also output 2 of the channels after every 4 coupling layers. The WaveGlow network was trained on 8 Nvidia GV100 GPU's using randomly chosen clips of 16,000 samples for 580,000 iterations using weight normalization~\cite{salimans2016weight} and the Adam optimizer~\cite{kingma2014adam}, with a batch size of 24 and a step size of $1\times 10^{-4}$  When training appeared to plateau, the learning rate was further reduced to $5\times 10^{-5}$.  

\subsection{Audio quality comparison}
We crowd-sourced Mean Opinion Score (MOS) tests on Amazon Mechanical Turk. Raters first had to pass a hearing test to be eligible.  Then they listened to an utterance, after which they rated pleasantness on a five-point scale. We used 40 volume normalized utterances disjoint from the training set for evaluation, and randomly chose the utterances for each subject. After completing the rating, each rater was excluded from further tests to avoid anchoring effects.  

The MOS scores are shown in Table~\ref{tab:mos} with 95\% confidence intervals.  Though MOS scores of synthesized samples are close on an absolute scale, none of the methods reach the MOS score of real audio.  Though WaveGlow has the highest MOS, all the methods have similar scores with only weakly significant differences after collecting approximately 1,000 samples.  This roughly matches our subjective qualitative assessment.  Samples of the test utterances can be found online~\cite{WaveGlowSite}.  The larger advantage of WaveGlow is in training simplicity and inference speed.

\begin{table}[!h]
\begin{center}
    \begin{tabular}{ p{3cm}|p{4.5cm} }
        \textbf{Model} & \textbf{Mean Opinion Score (MOS)}\\
        \hline
        Griffin-Lim        & $ 3.823 \pm 0.1349 $\\    
        WaveNet            & $ 3.885 \pm 0.1238 $\\
        WaveGlow           & $ 3.961 \pm 0.1343 $\\
        Ground Truth       & $ 4.274 \pm 0.1340 $\\
        \hline
    \end{tabular}
\end{center}
\caption{Mean Opinion Scores}
\label{tab:mos}
\end{table}

\subsection{Speed of inference comparison}
Our implementation of Griffin-Lim can synthesize speech at 507kHz for 60 iterations of the algorithm. Note that Griffin-Lim requires the full spectrogram rather than the reduced mel-spectrogram like the other vocoders in this comparison. The inference implementation of the WaveNet we compare against synthesizes speech at 0.11kHz, significantly slower than the real time.

Our unoptimized PyTorch implementation of WaveGlow synthesizes a 10 second utterance at approximately 520kHz on an NVIDIA V100 GPU.  This is slightly faster than the 500kHz reported by Parallel WaveNet~\cite{van2018parallel}, although they tested on an older GPU.  For shorter utterances, the speed per sample goes down because we have the same number of serial steps, but less audio produced. Similar effects should be seen for Griffin-Lim and Parallel WaveNet.
This speed could be increased with further optimization. Based on the arithmetic cost of computing WaveGlow, we estimate that the upper bound of a fully optimized implementation is approximately 2,000kHz on an Nvidia GV100.

%% file: discussion.tex
Existing neural network based approaches to speech synthesis fall into two groups.  The first group conditions future audio samples on previous samples in order to model long term dependencies. The first of these auto-regressive neural network models was WaveNet~\cite{van2016wavenet} which produced high quality audio.  However, WaveNet inference is challenging computationally.  Since then, several auto-regressive models have attempted to speed up inference while retaining quality~\cite{arik2017deep,arik2017deep2,jin2018fftnet}.  As of this writing, the fastest auto-regressive network is~\cite{kalchbrenner2018efficient}, which uses a variety of techniques to speed up an auto-regressive RNN.  Using customized GPU kernels,~\cite{kalchbrenner2018efficient} was able to produce audio at 240kHz on an Nvidia P100 GPU, making it the fastest auto-regressive model.

In the second group, Parallel WaveNet~\cite{van2018parallel} and \mbox{ClariNet}~\cite{ping2018clarinet} are discussed in Section~\ref{sec:introduction}. MCNN for spectrogram inversion~\cite{arik2018fast} produces audio using one multi-headed convolutional network. This network is capable of producing samples at over 5,000kHz, but their training procedure is complicated due to four hand-engineered losses, and it operates on the full spectrogram rather than a reduced mel-spectrogram or other features.  It is not clear how a non-generative approach like MCNN would generate realistic audio from a more under-specified representation like mel-spectrograms or linguistic features without some kind of additional sampling procedure to add information.

Flow-based models give us a tractable likelihood for a wide variety of generative modeling problems, by constraining the network to be invertible. 
We take the flow-based approach of~\cite{kingma2018glow} and include the architectural insights of WaveNet.  Parallel WaveNet and ClariNet use flow-based models as well.  The inverse auto-regressive flows used in Parallel WaveNet~\cite{van2018parallel} and ClariNet~\cite{ping2018clarinet} are capable of capturing strong long-term dependencies in one individual pass.  This is likely why Parallel WaveNet was structured with only 4 passes through the IAF, as opposed to the 12 steps of flow used by WaveGlow.  However, the resulting complexity of two networks and corresponding mode-collapse issues may not be worth it for all users. 

WaveGlow networks enable efficient speech synthesis with a simple model that is easy to train. We believe that this will help in the deployment of high quality audio synthesis.

%% file: acknowledgments.tex
The authors would like to thank Ryuichi Yamamoto, Brian Pharris, Marek Kolodziej, Andrew Gibiansky, Sercan Arik, Kainan Peng, Prafulla Dhariwal, and Durk Kingma.